# Microstructure-Dependent Particulate Filtration using Multifunctional Metallic Nanowire Foams

James Malloy,[1] Erin Marlowe,[1] Christopher J. Jensen,[1] Isaac S. Liu,[1,2] Thomas Hulse,[1,3] Anne F. Murray,[4] Daniel Bryan,[4] Thomas G. Denes,[4] Dustin A. Gilbert,[5] Gen Yin,[1] and Kai Liu[1,*]

[1]*Department of Physics, Georgetown University, Washington, DC 20057*

[2]*Department of Computer Science, Vanderbilt University, Nashville, TN 37235*

[3]*Department of Physics, University of Louisville, Louisville, KY 40292*

[4] *Department of Food Science, University of Tennessee, Knoxville, TN, 37996, USA*

[5]*Materials Science Department, University of Tennessee, Knoxville, TN 37996*

## Abstract

The COVID-19 pandemic has shown the urgent need for the development of efficient, durable, reusable and recyclable filtration media for the deep-submicron size range. Here we demonstrate a multifunctional filtration platform using porous metallic nanowire foams that are efficient, robust, antimicrobial, and reusable, with the potential to further guard against multiple hazards. We have investigated the foam microstructures, detailing how the growth parameters influence the overall surface area and characteristic feature size, as well as the effects of the microstructures on the filtration performance. Nanogranules deposited on the nanowires during electrodeposition are found to greatly increase the surface area, up to 20 m$^2$/g. Surprisingly, in the high surface area regime, the overall surface area gained from the nanogranules has little correlation with the improvement in capture efficiency. However, nanowire density and diameter play a significant role in the capture efficiency of PM$_{0.3}$ particles, as do the surface roughness of the nanowire fibers and their characteristic feature sizes. Antimicrobial tests on the Cu foams show a >99.9995% inactivation efficiency after contacting the foams for 30 seconds. These results demonstrate promising directions to achieve a highly efficient multifunctional filtration platform with optimized microstructures.

# 1. Introduction

Ultrasmall airborne particulates pose potentially severe health risks due to their ability to penetrate deep into the respiratory system, from the transmission of infections such as COVID-19 to air pollution.[1-4] One of the most effective methods of combatting the spread of COVID-19 has been the use of high efficacy face masks to prevent the inhalation of viral particles attached to ultrasmall aerosols and droplets.[5-7] A variety of materials have been used for particulate filtration, such as the cotton, fiberglass, and polypropylene. Cotton is inexpensive but has a low filtration efficiency;[2] fiberglass has a high filtration efficiency but is not structurally stable enough to be used as a face respirator;[8] and polypropylene (used in N-95 respirators) has a high filtration efficiency but cannot be easily cleaned, reused or recycled, due to the reliance on electrostatics. One of the major drawbacks of using disposable face masks has been the environmental toll,[9-11] with an estimated average 1.6 million metric tons of plastic being discarded every day during the pandemic.[12] It is critically important to develop reusable, efficient and robust filters for submicron airborne particulates, especially for protection against COVID-19 and other infectious diseases.

Many novel filter materials have been studied in light of the COVID-19 pandemic, including electrostatically charged Polyvinylidene fluoride (PVDF) nanofibers,[13] nanoparticle imbedded protein nanofibrils,[14] Ag-imbedded polymers,[15-17] porous metal–organic frameworks (MOFs),[18-21] and carbon nanotubes.[22-24] Metallic filtraton media with no polymer components are partcularly promising, as they can withstand high temperatures, ionizing radiation, and exposure to oils, all of which cause polymers to rapidly degrade.[25-27] Conventional porous metal foams often utilize sizable building blocks, with micron-scale dimensions and larger, which are inefficient against deep-submicron particulates that are the hardest to capture.[28-29] Recently advances in metallic nanowire networks and foams offer a promising new platform for explorations in catalysis, gas storage, nanomagnetics and neuromorphic computing.[30-35] However, most such metallic monoliths are not mechanically robust enough to be useful as practical filters. It is challenging to realize nanoporous metallic foams that maintain nanoscale dimensions, which are essential for deep micron particulate filtration, and are also mechanically robust.

Previously, we have demonstrated that nanowire-based low-density metal foams, synthesized by electrodeposition and sintering, are robust enough to withstand air speeds of over 20 m/s and pressure differentials of over several atmospheres without showing any sign of deformation or degradation.[36] Further, they also exhibit extremely large surface areas as well as outstanding filtration efficiencies (>96.6%) in the $PM_{0.3}$ regime with breathability comparable to N-95 respirators.[36] They can also be easily cleaned[36] and decontaminated[37] to allow sustained use, and eventually recycled at the end of their useful lifetime.[38]

Despite the promising initial results on such nanoporous metallic foams, a quantitative understanding of how the synthesis conditions and foam microstructures affect the filtration performance is still lacking. To enhance the foam mechanical stability, a second electrodeposition process was used to reinforce the intersections of the Cu foams.[36] An accompanying effect is a significant change in the foam surface areas and morphology due to nanogranule growths along the nanowires. In such 3D interconnected foams, it is critical to understand the correlation between the foam microstructures and the filtration performance.

In this study we have quantitatively investigated how the nanoporous foam microstructures impact the filtration performance, particularly against particulates that are around 0.3 μm in size used in standardized filtration efficiency tests.[28-29] We distinguish the effects of surface area enhancement vs. the density of nanoscale feature sizes comparable to the particulates being filtered. Interestingly, while foam surface areas are generally important for filtration efficiency, for high surface areas in the range of 1 m$^2$/g and beyond, the sizes of the foam surface features play a more prominent role. These results demonstrate nanoporous metallic nanowire foams as a multi-functional platform for deep submicron particulate filtration that is efficient, robust, antimicrobial, and reusable, with the potential to further guard against multiple hazards.

**Results and Discussion**

<u>Synthesis and foam morphology</u>: Nanoporous copper foams were fabricated using electrodeposited nanowires (NWs) following a cross-linking and freeze-drying technique, as described previously.[36,31] Anodized aluminum oxide (AAO) templates with 200 nm pore size and 60 μm thickness were used for the electrochemical deposition (ED) of Cu NWs.[34, 39-40] Subsequently, NWs were liberated from the AAO membrane into deionized water. The NW-water suspension was freeze-cast by liquid nitrogen into the desired shape and then pumped in vacuum to sublimate the ice. The resultant free-standing foam was strengthened by sintering at 300 °C, while simultaneously undergoing multiple oxidation / reduction cycles.[36, 41] Here, foams with an initial density $\rho_i$ ~2-3% of the Cu bulk density were achieved after this first synthesis step of ED and sintering. Such foams have extremely large surface area-to-volume ratios, up to $10^6 : 1$ m$^{-1}$.[31]

Subsequently, the foams were strengthened further using a secondary electrodeposition (2ED) step to achieve a final density $\rho_f$, in a bath of copper sulfate along with a mixture of leveler, accelerator, and suppressor compounds to ensure even and homogeneous plating.[36] The resulting foams were significantly stronger and more durable due to increased bonding areas at intersections of nanowires, while still retaining a high surface area and an excellent filtration efficiency.[36]

The morphology of the foams varied widely depending on the amount of secondary electroplating as well as other factors **(Fig. 1)**. As the 2ED process began, small nanogranules started to deposit along the surface of the NWs (Fig. 1a, for a $\rho_f$ =2.2% density foam). As the amount of copper plated increased, these nanogranules began to increase in size (Fig. 1b, for a $\rho_f$ =5% density foam). With enough plating, e.g., for a foam with a $\rho_f$=30% density, the nanogranules eventually coalesced together, enclosing multiple nanowires into larger filaments with smoother surfaces (Fig. 1c). The foam microstructures could be further tuned by the electrolyte conditions, e.g., lowering the pH value of the solution reduced the nanograin feature size deposited on the nanowire surface and increased the overall surface smoothness (see Supplementary Information). Final optimized foams exhibit both low density and strong mechanical strength that are essential for filtration media applications (**Fig. 2**ab).

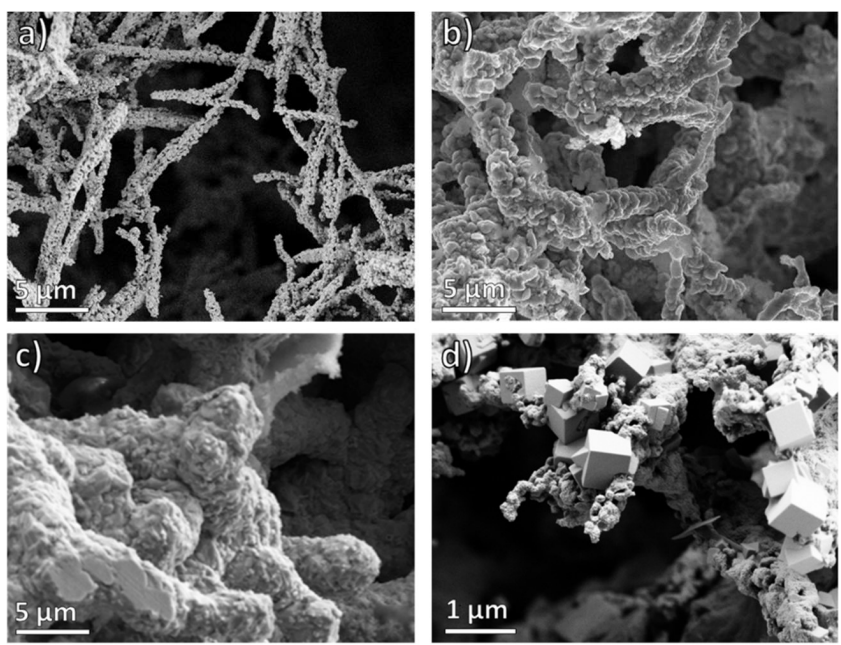

**Figure 1. Foam microstructure evolution with synthesis conditions.** SEM Images of 2ED (a) 2.2%, (b) 5%, and (c) 30% density foams, illustrating the evolution of foam morphology. With continued electrodeposition, the nanogranules along the nanowires grow larger, eventually coalesce into a thicker coating along the nanowires, even forming large filaments as shown in panel (c). (d) Cubical NaCl crystals captured in a 5% density foam.

Pressure differential and breathability: One of the most important characteristics of a filter is its breathability, which is inversely proportional to the pressure drop across the foam ($\Delta P$). The $\Delta P$ quantity is measured experimentally by flowing air through the foam as a function of air velocity and measuring the difference between the pressure upstream and downstream of the foam (Supplementary Information). It was shown previously[36] that $\Delta P = Av + Bv^2$, where $v$ is the velocity of the airflow, and A and B are constants determined by the properties and geometry of the foam.[42] As most filters operate in the low velocity regime where $v \approx 0.05$ m/s, the linear term is the dominant one,[43] or $\Delta P \approx Av$. Thus, it is more useful to evaluate the pressure drop coefficient $\Delta P/v$ for our foam samples with various $\rho_i$, $\rho_f$ and thickness ($t$), as shown in **Table 1** and **Fig. 2c.**

**Table 1: Foam Filtration Characteristics***

| Sample ID | $\rho_i$ (%) | $\rho_f$ (%) | $t$ (mm) | $\Delta P/v$ (kPa·m⁻¹s) | Efficiency (%) (Measured) | $F_{QE}$ | $F_{QC}$ | $F_{QC}$ Error (%) | $F_{QSF}$ | $Q_{Measured}$ | $Q_{Calcu.}$ |
|---|---|---|---|---|---|---|---|---|---|---|---|
| 1 | 2.0 | 3.5 | 0.9 | 0.48 ± 0.04 | 91.0 ± 0.1 | 2.41 ± 0.01 | 2.59 | 7.5 | 2.43 | 5.02 ± 0.41 | 4.87 |
| 2 | 2.0 | 5.0 | 0.9 | 0.58 ± 0.03 | 84.2 ± 3.0 | 1.85 ± 0.19 | 1.98 | 7.0 | 1.87 | 3.19 ± 0.36 | 3.47 |
| 3 | 2.5 | 5.4 | 1.0 | - | 92.5 ± 0.1 | 2.59 ± 0.01 | 2.92 | 13 | - | - | - |
| 4 | 3.2 | 5.0 | 0.9 | 1.88 ± 0.12 | 99.0 ± 0.1 | 4.61 ± 0.10 | 3.95 | -14 | 4.56 | 2.45 ± 0.15 | 2.68 |
| 5 | 2.5 | 6.8 | 1.0 | - | 89.1 ± 0.1 | 2.22 ± 0.09 | 2.60 | 17 | - | - | - |
| 6 | 2.0 | 9.0 | 1.0 | 0.73 ± 0.03 | 81.1 ± 0.5 | 1.67 ± 0.02 | 1.62 | -3.0 | 1.71 | 2.29 ± 0.10 | 2.31 |
| 7 | 2.0 | 15 | 1.0 | 0.97 ± 0.11 | 76.8 ± 1.3 | 1.46 ± 0.05 | 1.25 | -14 | 1.41 | 1.51 ± 0.18 | 1.22 |
| 8 | 2.0 | 15 | 2.5 | 2.76 ± 0.21 | 96.6 ± 2.0 | 3.38 ± 0.67 | 3.13 | -7.4 | - | - | - |
| 9 | 2.0 | 30 | 1.0 | 4.67 ± 0.24 | 54.2 ± 0.5 | 0.78 ± 0.01 | 0.88 | 13 | 0.77 | 0.17± 0.01 | 0.18 |

*Including $\frac{\Delta P}{v}$; measured $PM_{0.3}$ filtration efficiency; measured, calculated and surface-feature corrected filtration qualities ($F_{QE}$, $F_{QC}$ and $F_{QSF}$ respectively); and the measured and calculated quality factors Q.

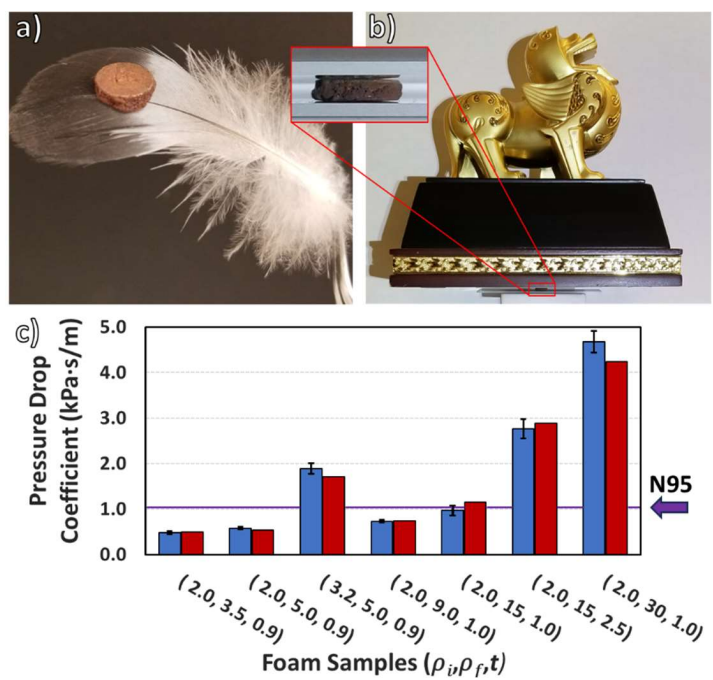

**Figure 2. Robust and breathable light-weight foam.** (a) A 120 mg 15% bulk density foam 9 mm in diameter resting on the tip of a feather. (b) The same foam supporting a 1.8 kg weight. (c) Comparison of measured (blue) vs. calculated (red) breathability for foam samples with various initial density $\rho_i$ (%), final density $\rho_f$ (%), and thickness $t$ (mm). A 15% bulk density foam (sample 7) similar to the one pictured in panels a,b has a comparable breathability to an N95 (purple line).

For comparison, an N-95 respirator has a pressure drop coefficient of around 1.0 kPa·m$^{-1}$s.[36] Our $\rho_f$ =15% foams (sample 7, $\Delta P/v$= 0.97± 0.11 kPa·m$^{-1}$s) were found to have breathability comparable to an N-95, while also being light-weight enough to rest comfortably on the tip of a feather (**Fig. 2a**) and structurally robust enough to support a 1.8 kg weight without collapse or deformation (**Fig. 2b**). Most of our 1 mm thick foams with $\rho_f \leq 15\%$ (samples 1-7) exhibit excellent breathability, which improves as foam thickness and packing density decrease, with the ultra-low density $\rho_f$=3.5% foam (Table 1, sample 1, ΔP/v= 0.48± 0.04 kPa·m$^{-1}$s) being over twice as breathable as an N-95 (ΔP/v= 1.02± 0.02 kPa·m$^{-1}$s). This overall trend is also illustrated in Fig. 2 (blue bins), which reveal a few trends: 1) An inverse proportionality between breathability and filter thickness is clearly seen for foams with the same initial and final densities (sample 7 vs. 8). 2) For foams with the same initial density $\rho_i$ and thickness (sample #1 vs. 2, #6 vs. 7 and 9), the higher the final density $\rho_f$, the larger the $\Delta P/v$, indicating that the additional Cu deposited during the 2ED process is generally impeding the air flow. 3) Interestingly, for foams with the same final density $\rho_f$ and thickness, sample 4 with $\rho_i$ of 3.2% has a $\Delta P/v$ that is more than 3 times that of sample 2 with $\rho_i$ of 2.0%. Both samples were made with the same starting 200 nm diameter Cu nanowires, however sample 4 was fabricated with a greater number density of nanowires which results in a significantly increased nanowire length density compared to sample 2. Further, the amount of Cu plated onto sample 4 per nanowire during the 2ED stage is significantly less than in sample 2, which results in smaller final diameter nanowires. Both the increased nanowire length density and the decreased nanowire diameter in sample 4 contribute to a higher pressure drop due to the increased surface area per filter volume which induces more drag.[44]

To gain a quantitative understanding of the exact relationship between breathability and nanowire diameter and foam initial/final densities, as a crude first approximation, we may consider the foams as a simple compilation of isolated cylindrical fibers with smooth surfaces, even though the actual foams are more complex interconnected 3D nanowire networks with various surface details. Single fiber theory for the pressure drop[44], combined with the Miyagi model[45] for the pressure drop across an array of fibers allows us to express the linear pressure drop coefficient as (derivation in Supplementary Information):

$$\frac{\Delta P}{v} = -C \frac{\rho_i^{\frac{5}{2}} \cdot e^{4.75\rho_f} \cdot t}{\rho_f^{\frac{1}{2}} \cdot \left[\ln(\rho_f) + \frac{1-\rho_f^2}{1+\rho_f^2}\right]} \quad (1)$$

Remarkably, with this substantial simplification, the calculated pressure drop is quite comparable to the measured results **(Fig. 2c)**. This agreement suggests that pressure drop scales linearly with filter thickness ($t$) and increases substantially as the initial nanowire number density ($\propto \rho_i$) and final density ($\rho_f$) go up. By fitting Eqn. 1 across a range of foam samples, the scaling factor $C$ is determined to be $3.64 \times 10^6$ kPa·m$^{-2}$s.

Filtration efficiency: Foam samples of varying thicknesses, $\rho_i$ and $\rho_f$ are used to filter aerosolized 0.3 μm NaCl particles, and the overall filtration efficiency is measured using a TSI AeroTrak® Portable Particle Counter Model 9110 (Supplementary Information). A few clear trends can be seen between samples that only differ by one parameter (**Table 1**). For instance, for foams with the same microstructures (e.g., samples 7 and 8), a thicker filter is indeed more efficient; for foams with the same $\rho_i$ and $t$, smaller $\rho_f$ leads to a higher filtration efficiency (samples 6, 7, and 9). Interestingly, for foams with the same $\rho_f$ and $t$, sample 4 with a higher $\rho_i$ of 3.2 % exhibits a higher filtration efficiency of 99.0%, as compared to sample 2 ($\rho_i$ =2.0 %, with an 84.2% efficiency). While the correlation between filter thickness with filtration efficiency is well understood, the relationship between the foam density and other key microstructure characteristics with filtration efficiency is less clear.

To evaluate the correlation between the foam microstructure and filtration efficiency, we will first model the capture efficiency using basic air filtration theory for a perfectly smooth fiber. The filtration quality of a material is defined as a dimensionless quantity $F_Q \equiv -\log(1-E)$, where $E$ is the filtration efficiency.[2, 44] An N-95 respirator is certified to have a $F_Q$ of at least 3.0 for particles in every size range, while an N-99 respirator has a $F_Q$ of at least 4.6.[46] Using the smooth fiber model,[3, 44] the filtration quality can be written as (see Supplementary Information):

$$F_Q \sim \frac{n*t}{d_f} = k_E \frac{\rho_i^{\frac{3}{2}} \cdot t}{\rho_f^{\frac{1}{2}}} \qquad (2)$$

where $n$ is the length density of the nanowires in the foam, or the sum of the lengths of all the nanowires per given volume, and it scales with $\rho_i$; $d_f$ is the final average nanowire diameter after electrodeposition and scales as $d_f \propto \sqrt{\frac{\rho_f}{\rho_i}}$, and finally $k_E$ is a constant dependent on a multitude of factors such as filtration material, fiber geometry, and other testing conditions that affect filtration efficiency. As it is more useful to evaluate how far off proportionally the calculated filtration quality $(F_{QC})$ is from experimentally measured filtration quality $(F_{QE})$, rather than the difference between the two values, we fit $k_E$ to minimize $\sum \left( \frac{F_{QC}}{F_{QE}} - \frac{F_{QE}}{F_{QC}} \right)^2$. This ensures that foams with larger thicknesses or otherwise have higher filtration qualities are not weighted disproportionately, and that the percentage error between $F_{QE}$ and $F_{QC}$ are minimized with equal weight across all ranges of foam samples with varying filtration qualities. The best fit for the smooth fiber model described in equation 2 was determined to be $k_E = 171.5 \ mm^{-1}$.

As shown in Table 1, the smooth fiber model alone can be used to describe the filtration quality of the foam filters to within an error of ±17%. With a theoretical baseline for how well perfectly smooth fibers capture particles, we can now determine *to what extent surface features contribute towards particulate filtration*. It is well established that a rougher fiber surface improves filtration capture efficiency,[47-49] which is intimately coupled with an increase in surface area and the foam microstructure. Initially the nanowires which comprise the foam have a relatively smooth surface structure as they were grown in an anodized aluminum oxide mold. However, during the 2ED process the deposition of Cu nanogranules on the surface of the nanowires begin to dramatically increase the surface area. Surface area was measured to high accuracy (< 0.5% error) with a BET krypton adsorption isotherm. It was found that foams that were initially ~3% bulk density pre-2ED exhibit a maximal surface area of 20 m²/g for a 13% bulk density after 2ED. The surface area was found to initially increase as the relative density increases, with the percentage density increase being measured in coulombs of copper deposited during the 2ED process per mg of 1ED foam **(Fig. 3a)**. In foams with an even higher density, the deposited nanogranules begin to coalesce into a smoother surface structure and the surface area decreases.

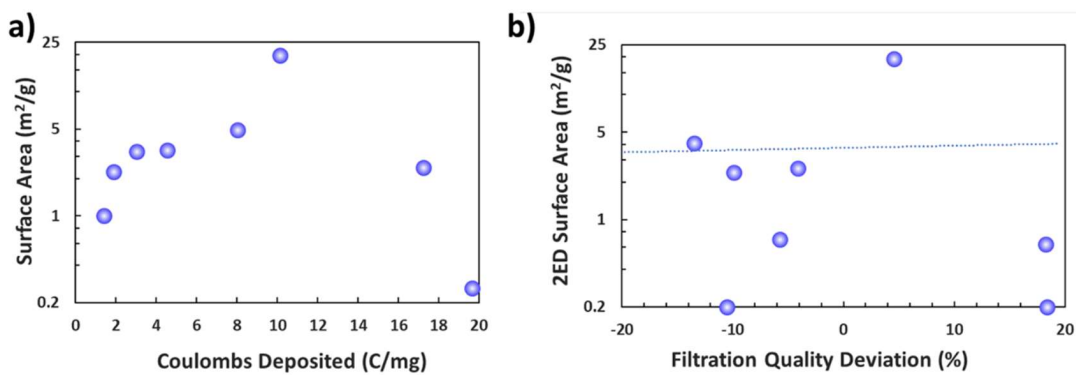

**Figure 3. Surface area correlation with filtration performance.** (a) Surface area of the foams versus the relative amount of copper deposited during the 2ED step (coulombs deposited per mg of 1ED foam). Surface area begins to increase as deposited nanogranules along the nanowire surface increase overall surface roughness, up to around 20 m²/g. Somewhere after this point, nanogranules begin to collesce into a smoother surface structure with reduced surface area. (b) Filtration quality deviation showing no obvious correlation ($r^2 = 2\times10^{-5}$) with the increased surface area contributed from deposited surface features.

Previously we have observed a substantial drop in filtration efficiency in the $\rho_f$ = 30% foam, whose surface area is more than an order of magnitude smaller than other foams with a lower $\rho_f$. It is interesting to examine any direct correlation between surface area and filtration efficiency. With the 2ED process, it was found that the deposited surface features increase the internal surface area of the foam substantially, some by over a factor of 10 higher than what is predicted by the smooth fiber model, e.g., the aforementioned sample with a 20 m²/g surface area. To determine how much the increased surface area from the foam's surface features might influence capture efficiency, additional surface area from 2ED is compared with the smooth fiber filtration quality model, as illustrated in **Fig. 3b**.

Correlation with foam surface area and microstructure: Interestingly, for foams with surface area in the > 1 m²/g range, the additional surface area contribution from 2ED has no appreciable correlation with the capture efficiency of 300 nm sized particles, as a linear regression model yields a $r^2$ value of $2\times10^{-5}$. This suggests that surface roughness and the surface area it adds as a stand-alone metric does not inherently aid in particle capture efficiency. But rather, there must be an additional characteristic of surface roughness that is required for filtration efficiency improvement. It has been shown on flat 2D surfaces that particles are captured more easily when they can nest into similarly sized crevices which increase the area of contact between the particle and substrate, leading to a higher sticking coefficient.[50] This can be seen in Fig. 1d, where the NaCl crystals tend to be wedged into comparably sized crevices which gives a higher area of contact. Therefore, we want to test how crucial of a role the *size* of the features deposited onto the fiber surface plays in influencing filtration efficiency.

Image analysis on SEM micrographs was performed on foam samples of various densities to determine the feature size distribution across samples. After preprocessing to remove blurry background, an auto-local-threshold filter is applied to extract and measure the size and frequency of the nanogranule deposits along the nanowire surface (more details in Supplementary

Information). From this we can determine the feature size distribution across various samples, as illustrated in **Fig. 4**.

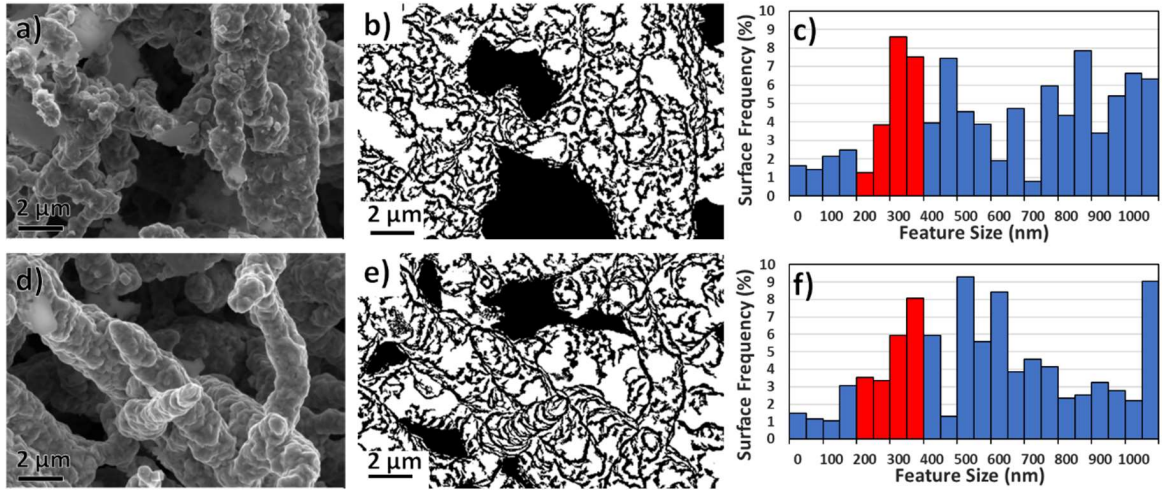

**Figure 4. Foam nanogranule size analysis.** Image analysis performed on interiors of (a-c) $_f\rho$=9% and (d-f) 17% foams, showing (a,d) SEM images, (b,e) images with thresholding, and (c,f) feature size distribution histograms with features between 200-400 nm highlighted in red.

The increased particle capture due to nesting is maximized when surface features along the nanowires are similar in size to the particles being captured, and becomes diminished as the surface features become too large or too small relative to the particles. While the theoretical smooth fiber model provides a good fit to the experimental results **(Fig. 5a)**, there are still appreciable differences due to surface features. To determine what the capture efficiency contribution is from the surface features, we can cross-reference the percentage of surface features in the 200-400 nm size range **(Fig. 5b)** with the filtration quality deviation between our measured results and the smooth fiber model (**Fig. 5a**).

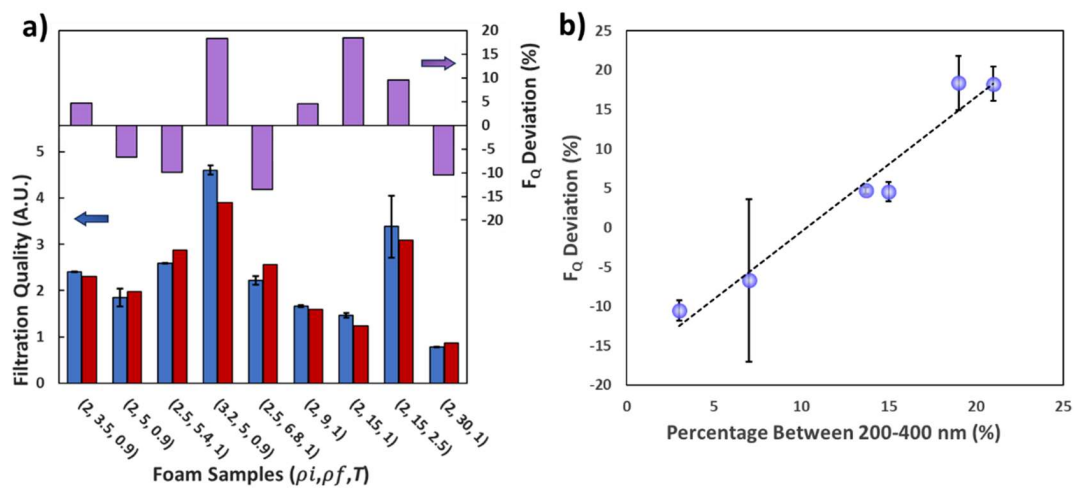

**Figure 5. Filtration quality dependence on foam microstructures**. (a) Measured (blue) versus calculated (red) filtration quality histogram and filtration quality deviation (purple) between measured and calculated values. (b) High correlation ($r^2 = 0.96$) between filtration quality deviation and percentage of surface features between 200-400 nm.

By directly comparing the deviation in filtration quality for 300 nm sized particles between our measured results and the calculated smooth fiber model to the percentage of surface area that has surface features between 200-400 nm in size, we find that there is a very high correlation with the $r^2$ value of 0.96 in a simple linear regression model, as shown in Fig. 5b. This suggests that the size of the surface structure plays a prominent role in increasing filtration efficiency, with some samples showing over a 30% increase in filtration quality compared to smooth fibers. Furthermore, the filtration efficiency dependence on feature size provides an explanation for why rough surfaces and high surface area do not necessarily result in an increase to capture efficiency. This can be understood as much of the added surface area from the 2ED process is primarily from ultrasmall surface features that are simply too small relative to the 300 nm sized particles to enhance particle capture efficiency via nesting, and also too small to change the geometry of the fiber to increase capture through diffusion or interception.

By approximating a linear relationship with the percentage increase in filtration quality and percentage of surface structures in the 200-400 nm range, a surface feature adjusted filtration quality can be derived as

$$F_{QSF} = 142 \cdot \frac{\rho_i^{\frac{3}{2}} \cdot t}{\rho_f^{\frac{1}{2}}} (1 + 0.019P) \tag{3}$$

where $P$ is the percentage of surface area that has 200-400 nm sized surface features, and $t$ in mm. This equation yields an excellent fit with our measured results, with a linear regression model $r^2$ value of 0.999. Using both Eqn. 3 and Eqn. 1, we may now calculate the quality factor (Q) of the foams as

$$Q \equiv \frac{v \cdot F_{QSF}}{\Delta P} = -.039 \frac{\left[ \ln(\rho_f) + \frac{1 - \rho_f^2}{1 + \rho_f^2} \right] \cdot (1 + 0.019P)}{\rho_i \cdot e^{4.75\rho_f}} \tag{4}$$

This allows us to optimize the quality factor of the foam by tuning the initial and final densities of the foam. This can be seen in **Table 1** where an ultra-low density 3.5% density foam (Sample 1) was able to achieve a quality factor of 5.02, over 50% better than a standard N-95 respirator which has a quality factor of around 3.20.[36]

Note that our study has mostly focused on studying the effects of the nanowire length density, plating ratio during the 2ED stage, and surface features. There are a number of other parameters that can also influence both breathability as well as filtration efficiency that remained constant throughout the range of samples measured. As such, their contributions to filtration efficiency and breathability could be absorbed into the fitted proportionality constant. For instance, the spacing geometry of the NWs is important. When the NWs bundle together such that they create large pockets absent of NWs, the breathability increases dramatically. However, this benefit is offset by the dramatic drop in filtration efficiency. Similarly, the studies were carried out on samples that were in the density range of being structurally stable and NWs being relatively inflexible. While the derived equations would suggest that ultra-low-density foams would have an extremely high breathability and quality factor, it was found experimentally that at densities below 3% the NWs begin to collapse, limiting both breathability and filtration efficiency. Additionally, while the derived quality factor equation suggests that extremely high-quality factors can be achieved by reducing the NW density, realistically at these limits, the heavily diminished structural integrity and changes to NW spacing begins to play a non-negligible role which causes the quality factor to decrease. Additionally, the choice of NW materials may affect both the sticking coefficient of airborne particles and the rigidity of the foams, which are relevant to the pressure drop as well as filtration quality, which is of interest for future studies.

Another important functionality of the foams is its antimicrobial characteristics. Once captured, conventional filter materials sequester the live pathogens, which can survive in the respirator. This makes the used filter material a potential hazard as the pathogens can be released from the filter through rough handling.[51] Copper, being a bioactive metal with antimicrobial properties achieved through surface contact,[52] is expected to be self-sterilizing, killing captured bacteria and inactivating captured viruses.[15, 53] Antimicrobial testing was performed on our Cu foams using a contact assay described previously.[52] In these tests, 10 μL of a concentrated solution containing the bacteriophage Phi6 was deposited on the surface of the foam, then a glass slide was set on top of the droplet to disperse the solution across a 6.4 cm$^2$ area. After 30 seconds, the slide was removed and the solution was washed from the foam using 990 μL of phosphate buffer solution, which was collected in a petri dish. The solution was then enumerated by double-layer agar plating using its host bacteria *Pseudomonas syringae*. The bacteriophage Phi6 has a similar structure to SARS-CoV 2, including a double capsid lipid-protein envelope and protruding spike proteins. Bioactivity from a stainless steel coupon was also measured as a negative control. Each test was performed in triplicate.

The copper foam resulted in a 6.7 ± 1.3 log-reduction of the active bacteriophage, corresponding to a >99.9995% inactivation efficiency in the 30 second test. These results are consistent with the strong bioactivity of copper and demonstrate that this bioactivity extends to the nanostructured material. Functionally, these results confirm that captured pathogens will become inactivated soon after contacting the Cu superstructure.

**Conclusion**

We have demonstrated nanoporous metallic nanowire foams as a smart, multifunctional platform for deep submicron particulate filtration that is efficient, robust, antimicrobial, and reusable, with the potential to further guard against multiple hazards. By tuning the 2ED process to maximize the amount of deposited nanogranules and surface area, foams with surface areas of 20 $m^2$/g have been achieved, with breathability comparable to an N-95 respirator. Furthermore, we have shown how foam microstructures, including NW length density and the 2ED surface feature distribution, influence the capture efficiency. We have shown that overall surface roughness and its contribution to surface area is only a partial requirement for increasing filtration efficiency, and that the size of the surface features plays a pivitol role as well. This suggests that our foams, as well as other conventional filters, can be further improved by optimizing the size scale of the surface roughness of the fibers to capture 0.3 μm sized particles. By adjusting growth parameters and increasing the amount of ~0.3 μm sized surface features, the capture efficiency of a 0.9 mm thick 5% density foam is increased from 84% to over 99% for the $PM_{0.3}$ regime, and the filtration quality factor can be enhanced by over 30%, with room for further increase. Antimicrobial tests on the Cu foams demonstrate a >99.9995% inactivation efficiency after contacting the foams for 30 seconds. These results demonstrate promising directions to achieve a highly efficient multifunctional filtration platform with optimized microstructures.

**Supplementary Information**

Supplementary Information is available from the author.

**Acknowledgements**

This work has been supported by the Earth Commons Impact Awards, OTC Gap Funds, and the McDevitt bequest at Georgetown University. T.H. has been supported by the NSF-REU program (DMR-1950502 ).

**Author contributions**

J.M. and K.L. conceived the concept and K.L. coordinated the project. J.M. led the sample synthesis, microstructure characterizations, filtration measurements and data analysis, together with E.M., C.J.J., I.S.L., T.H., G.Y. and K.L.; I.S.L. and G.Y. performed imaging analysis. A.F.M., D.B., T.G.D. and D.A.G. performed antimicrobial tests. J.M. and K.L. wrote the first draft of the manuscript. All authors contributed to the revision.

**Conflict of interest statement**

J.M. and K.L. are coinventors on a filed patent application by Georgetown University on *Nanoporous metal foam gas and fluid filters*.

**Supplementary Information**

**Microstructure-Dependent Particulate Filtration Performance of Multifunctional Metallic Nanowire Foams**

James Malloy,[1] Erin Marlowe,[1] Christopher J. Jensen,[1] Isaac S. Liu,[1,2] Thomas Hulse,[1,3] Anne F. Murray,[4] Daniel Bryan,[4] Thomas G. Denes,[4] Dustin A. Gilbert,[5] Gen Yin,[1] and Kai Liu[1,*]

[1]Department of Physics, Georgetown University, Washington, DC 20057

[2]Department of Computer Science, Vanderbilt University, Nashville, TN 37235

[3]Department of Physics, University of Louisville, Louisville, KY 40292

[4] Department of Food Science, University of Tennessee, Knoxville, TN, 37996, USA

[5]Materials Science Department, University of Tennessee, Knoxville, TN 37996

## 1. Methods

### 1.1 Isotherm Measurements

Foam surface areas were determined using BET krypton adsorption isotherm measurements with error below 0.5%, carried out by the Particle Testing Authority of Micromeritics Instrument Corp. All samples were placed under vacuum at 200°C for 16 hours to properly degas before measurement.

### 1.2 Foam Feature Size Analysis

Size analysis of the foam surface features resulted from the 2ED process was performed for a variety of samples by first taking scanning electron microscopy (SEM) images of the foams, and then image-conditioning them using ImageJ to quantify the microstructures using developed algorithms in Python scripts. A flowchart of the full process is shown in Figure S1.

**Supplementary Information**

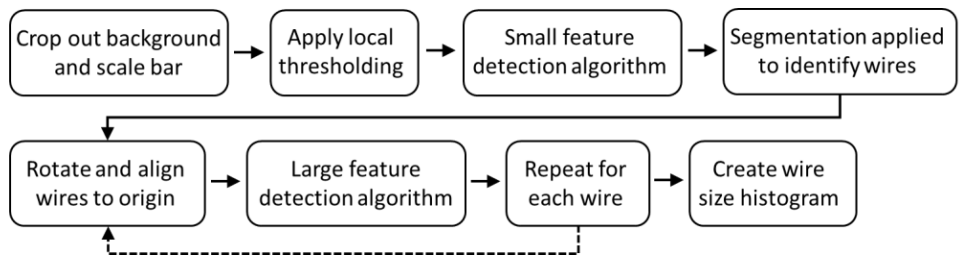

**Figure S1:** Image analysis flowchart.

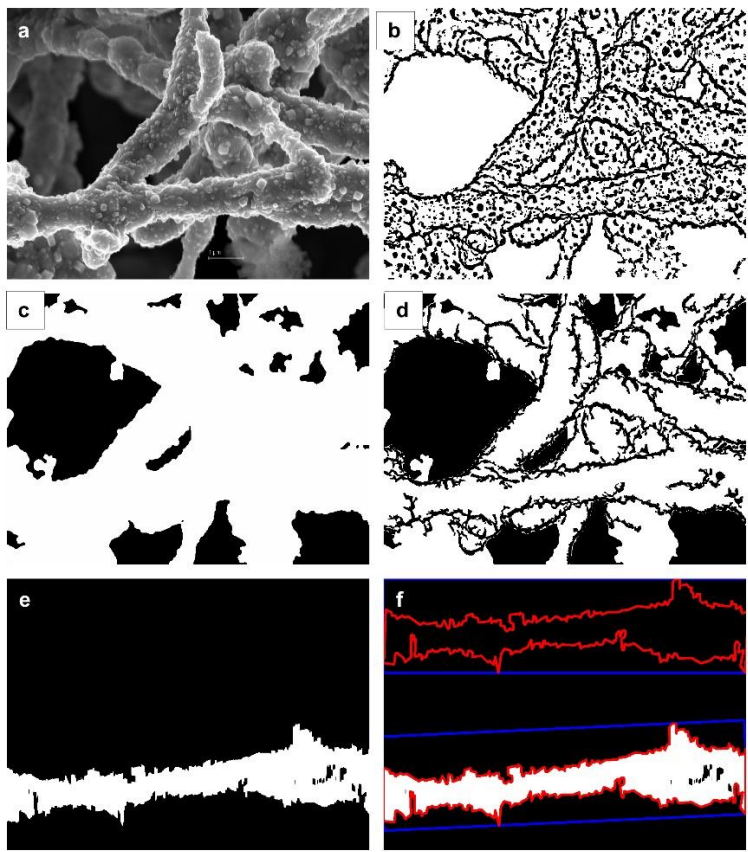

**Figure S2**: a) Initial image from SEM. b) Image after median auto-local-thresholding. c) Result of image segmentation. d) Large feature boundaries superimposed on segmented image. e) Single wire area extracted by large feature detection algorithm. f) Bounding box and axis-aligned wire outline used to calculate diameter.

**Supplementary Information**

The SEM images were prepared for analysis by removing the background or any unfocused areas and then removing scale legends and other irrelevant objects so that only features of interest remained. The analysis process began with small-scale feature detection. Separating small features such as nanogranules from 2ED from larger ones such as the nanowires was necessary because the two kinds appeared distinct. Small features appeared in the thresholded image as filled ellipses, whereas large ones would only have their outlines visible and appeared as long trunk-like features.

An *ImageJ* Auto Local Threshold filter using the median method was applied to the image to binarize the image and reveal all features (Fig. S2b). Next, the small feature detection algorithm was run on the binary image. This algorithm found small ellipse features with an area smaller than a threshold amount (such as 1000 pixels) and measured their approximate diameter. Subsequently, the identified small features were removed from the binary image so that only large features remained. The original image was then run through a segmentation filter in *ImageJ* using the *DiameterJ* plugin (Fig. S2c). The outlines of the large features were superimposed to divide the area into individual wires (Fig. S2d). Each wire was then analyzed individually (Fig. S2e). The wire was rotated so that its bounding box is horizontal and the diameter was measured at each pixel along the wire (Fig. S2f). The measured diameters were then averaged for the entire nanowire. This diameter quantification process was repeatedly applied to each nanowire in the input image.

Finally, a feature size distribution histogram was calculated. For every pixel in a feature, the feature diameter was added to the histogram, yielding a correlation between feature diameter and the number of pixels in the image representing features of the corresponding size. The size distribution histograms (Fig. S3), with features between 200-400 nm highlighted in red, can then

**Supplementary Information**

be used to determine the feature size evolution of the foams as a function of the 2ED process, which is included in Fig. 4 of the main text.

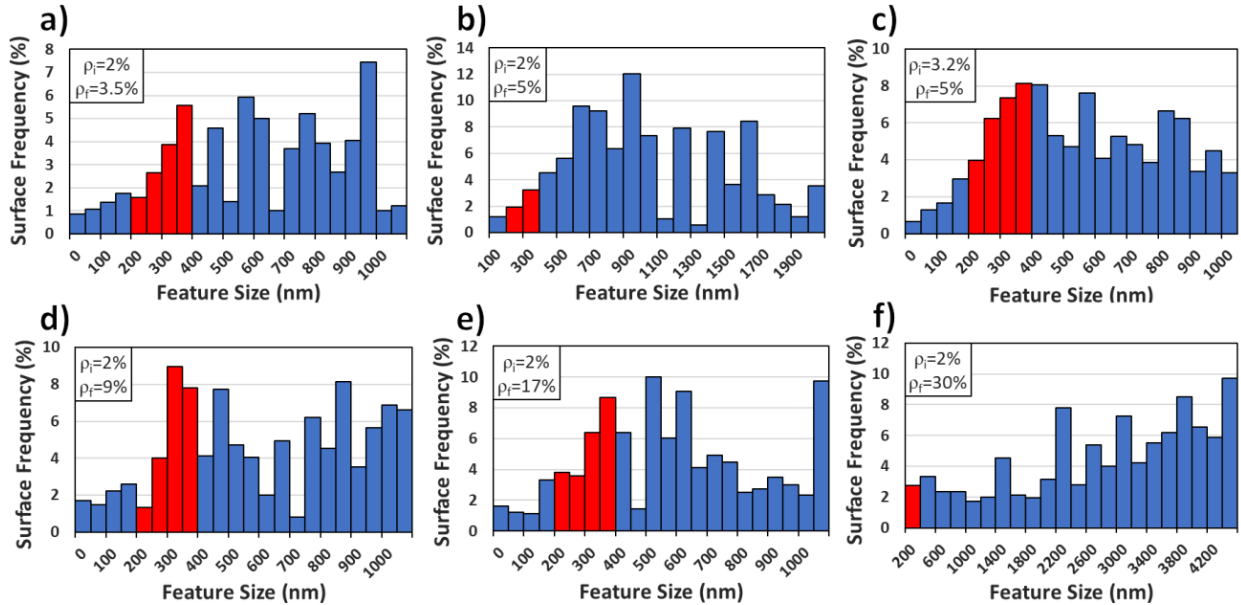

**Figure S3**: Feature size distribution of foam samples with (a) $\rho_i=2\%$, $\rho_f=3.5\%$, (b) $\rho_i=2\%$, $\rho_f=5\%$, (c) $\rho_i=3.2\%$, $\rho_f=5\%$, (d) $\rho_i=2\%$, $\rho_f=9\%$, (e) $\rho_i=2\%$, $\rho_f=17\%$, and (f) $\rho_i=2\%$, $\rho_f=30\%$.

**1.3 Pressure Drop Measurement**

Pressure differentials between the up-stream/down-stream of the foams were measured while the air flow rate was monitored with a Kelly Pneumatics Air & Oxygen Mass Flow Meter. The face velocity, or the speed at which air flows through the filter (air speeds are between 0 to 1 m/s), was determined by dividing the flow rate by the surface area of the foam. The pressure drop was then measured as a function of face velocities.

**Supplementary Information**

To determine how key characteristics of the foam influence the breathability, we first model the foam using the single fiber model for the pressure drop.[1] Air filtration theory allows expression of the linear pressure drop coefficient ($\frac{\Delta P}{v}$) as:

$$\frac{\Delta P}{v} = C_{\Delta P} \frac{4\eta tn \cdot f(\rho_f) \cdot f(K_n)}{d_f^2} \tag{S1}$$

where $\eta$ is the coefficient of viscosity, $t$ is the filter thickness, $f(\rho_f)$ is a function dependent on the packing density of the material, $n$ is the nanowire length density, $d_f$ is the nanowire diameter after plating,[1] and $C_{\Delta P}$ is a constant dependent on nanowire spacing geometry and various other factors. The mean free path of the gas molecules (λ) is around 67 nm at standard pressure and temperature. As λ is comparable to the many nanogranular growths along the nanowires as well as the diameter of the nanowires themselves, a significant amount of slip occurs which increases the drag force and is represented by a function of the Knudsen number ($K_n = \frac{2\lambda}{d_f}$).[2] The resulting pressure drop of the foams is found to scale as

$$\frac{\Delta P}{v} \propto \frac{nt \cdot f(\rho_f)}{d_f^3} \tag{S2}$$

The $f(\rho_f)$ function used is dependent on the geometry of the nanowires, which for the foams we can use the Miyagi cell model[3] written to a close approximation as:

$$f(\rho_f) \approx \frac{-8 \cdot \rho_f \cdot e^{4.75 \rho_f}}{\ln(\rho_f) + \frac{1 - \rho_f^2}{1 + \rho_f^2}} \tag{S3}$$

The average diameter of the nanowire before and after the 2ED process increases approximately as $d_f = d_i \sqrt{\frac{\rho_f}{\rho_i}}$, where $d_i$ is nanowire diameter before plating, $\rho_i$ and $\rho_f$ are the initial and final

**Supplementary Information**

foam density before and after 2ED, respectively. This can be derived from the volume of a cylinder equation, where $\rho_f \propto \frac{\pi}{4} n d_f^2$ and $\rho_i \propto \frac{\pi}{4} n d_i^2$. However, as all the nanowires in this study have the same initial diameter prior to plating, this can be simplified to $d_f \propto \sqrt{\frac{\rho_f}{\rho_i}}$. Additionally, the nanowire length density of the foam, which is the sum of the lengths of all the nanowires per given volume, scales as $n \propto \rho_i$, as it is proportional to the number density of nanowires. Combining those into Equations S2 and S3, we find:

$$\frac{\Delta P}{v} \propto - \frac{\rho_i^{\frac{5}{2}} \cdot e^{4.75 \rho_f} \cdot t}{\rho_f^{\frac{1}{2}} \cdot \left[\ln(\rho_f) + \frac{1 - \rho_f^2}{1 + \rho_f^2}\right]} \qquad (S4)$$

### 1.4 Foam Efficiency Measurement

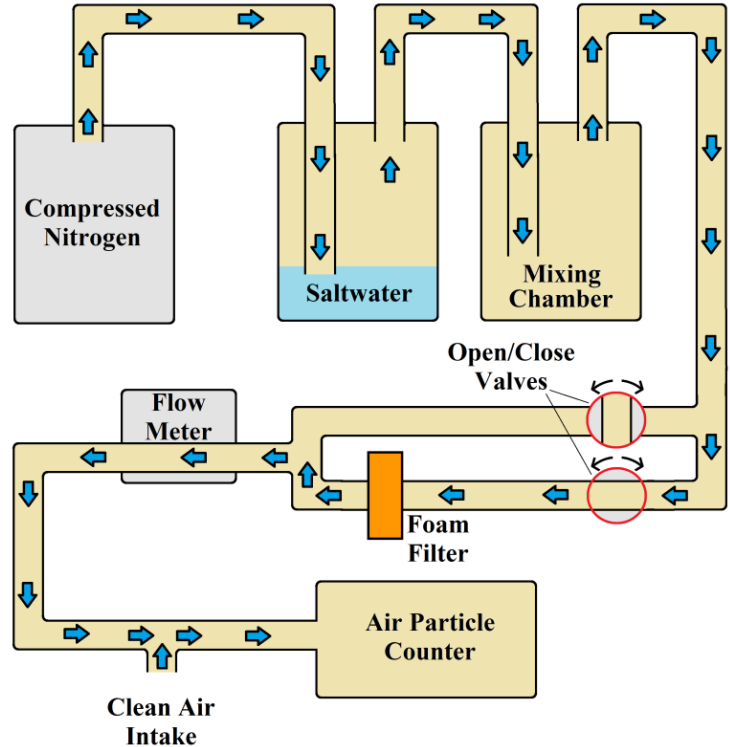

**Figure S4:** Foam efficiency measurement schematic.

**Supplementary Information**

Filtration efficiency was measured using a TSI 9110 portable particle counter with the setup depicted below (**Figure S4**). Compressed nitrogen was first flown through a container of stirred saltwater to generate NaCl particles. The NaCl particles were then flown into a large mixing chamber reservoir to keep the particle concentration stable, after which the particles entered into a junction where the open/close valves were positioned to either flow the air through the foam filter, or through an unobstructed path. The flow meter was used to measure the volume of the particle-filled air passing through, and an injection of clean air at atmospheric pressure through the clean air intake was used to bring the air pressure back up to atmospheric pressure and raise the total flow rate passing through the particle counter to 28.3L/min, both of which were required for the particle counter to function properly. As the particle counter was measuring the concentration of particles only after being mixed with clean air, the particle concentration before being mixed was calculated using

$$Q_m \cdot n_f + (Q_T - Q_m) \cdot n_c = Q_T \cdot n_m, \tag{S5}$$

where $Q_m$ was the air flow rate as measured by the flow meter, $n_f$ was the particle concentration flowing past the filter, $Q_T$ was the total air flow passing through the particle detector (28.3L/min), $n_c$ was the particle concentration of the clean air, and $n_m$ was the particle concentration as measured by the particle counter. Solving for $n_f$, we got:

$$n_f = \frac{Q_T}{Q_m} \cdot (n_m - n_c) + n_c \tag{S6}$$

The efficiency of the filter (E) could then be found by comparing the particle concentration when flowing NaCl particles past the filter, to the particle concentration when unfiltered:

$$E = 1 - [\frac{n_{f(filter)}}{n_{f(no\ filter)}}] \tag{S7}$$

**Supplementary Information**

This measurement was repeated several times for each sample to ensure concentration of generated NaCl particles remained stable. The pressure from the compressed nitrogen was used to adjust the air flow through the filter, with measurements typically performed at a flow rate of 0.5-1L/min.

## 2. Characterizations

### 2.1 Dependence of foam morphology on electrolyte pH

The pH value of the electroplating solution is important as more acidic solutions will start to etch away the foam during the plating process. This can be used to control the morphology of the foam as the etching is non-uniform. One notable difference is that the amount of copper dissolved in a specific area is correlated with high surface area, which results in acidic plating solutions preferentially targeting the deposited nanogranules and creating foams with smoother nanowires and less overall surface area.

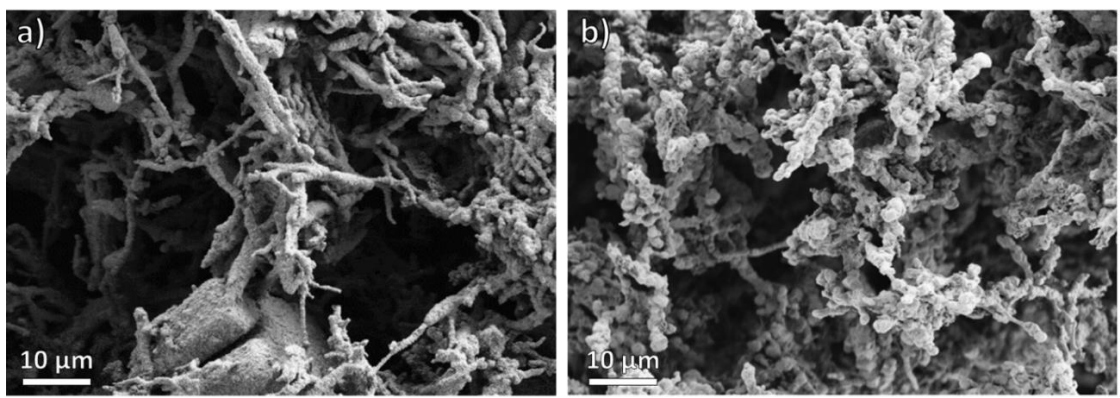

**Figure S5.** SEM image of (a) exterior surface and (b) interior region of a $\rho_f$ = 9% foam plated at pH of 1-3.

The benefit of this is that the pH can be tuned to maximize the percentage of nanogranules in the 0.3 μm size range. Various foam samples were plated with an electrolyte pH ranging between

**Supplementary Information**

1 to 4. The electrolyte becomes more acidic as it is used in multiple depositions. Foam plated with lower pH (≤ 3) electrolyte solutions were found to have smaller diameter nanowires along the exterior and larger diameter nanowires along the interior. The reason for this is that as the electrolyte solution diffuses towards the interior of the foam, it will etch away copper along the way and raise the pH locally as the solution becomes saturated with copper ions. This causes the plating solution in the interior of the foam to become locally less acidic than the solution in the exterior which results in the exterior nanowires being etched at a higher rate and having a disproportionately smaller diameter, as shown in Figure S5.

In cases where the foams were plated with an electrolyte on the higher end of the tested pH range (>3), the reverse trend was observed. Large nanogranules and large diameter nanowires are observed on the exterior of the foam while smooth, narrow nanowires are found in the interior. This is due to the 2ED process preferentially plating the exterior of the foam as result of a larger accumulation of charges. In lower pH solutions this 2ED growth was offset by the acidity of the electrolyte preferentially etching away at the exterior, while in higher pH solutions the 2ED growth prevails. Figure S6 shows a foam electrodeposited at a pH of 3.42, which has large nanogranules on the exterior surface. As the exterior is not being rapidly etched by the electrolyte, copper is able to nucleate on the nanowires. As the granules on the exterior of the foam grow, the electrolyte becomes less saturated with $Cu^{2+}$, lowering the pH inside the foam. This causes a similar effect to etching of the exterior of the foams plated at lower pH, resulting in narrower, smoother nanowires.

**Supplementary Information**

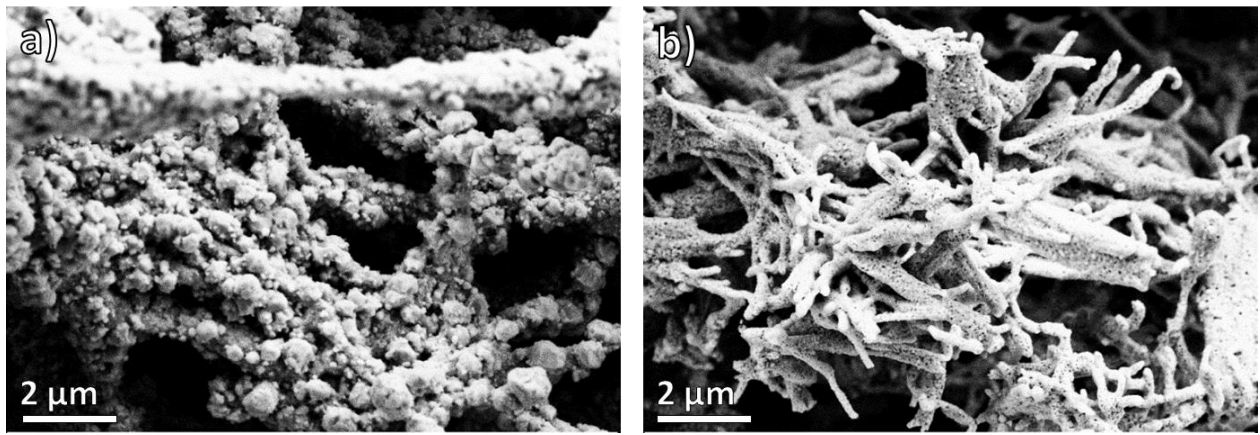

**Figure S6.** SEM image of (a) exterior surface and (b) interior region of a $\rho_f$ = 9% foam plated at pH of 3.42.

**2.2 Quantitative Foam Filtration Characteristics**

The calculated surface feature adjusted Filtration Quality ($F_{QSF}$) was found to be in very close agreement with the measured results, with $F_{QSF}$ generally being within the measurement error range of $F_{QE}$, as seen in **Table** 1 of the main text. The calculated Quality Factor $Q_{Calculated}$ was also found to be in close agreement with the measured results as well, and this remained true across a wide range of samples. The only sample where the calculated Quality factor is 2 sigma outside the error range of the measured quality factor is in Sample 7, where the likely cause is due to the heavily enhanced surface area as well as altered surface microstructure having a significant effect on the breathability. Any breathability dependence on surface feature microstructure results in the calculated Quality Factor not having as close of a match to the measured results as $F_{QSF}$ does, as the effects of the surface area and surface features on breathability were not examined in this study.

**Supplementary Information**

**2.3 Size Tunability and prototyping**

While the foams presented in this study are typically between 5-10 mm in diameter, we have demonstrated that there is no significant limitation on the size that can be manufactured. Pictured below in **Figure S7** is a 1 mm thick foam that is 4 cm in diameter.

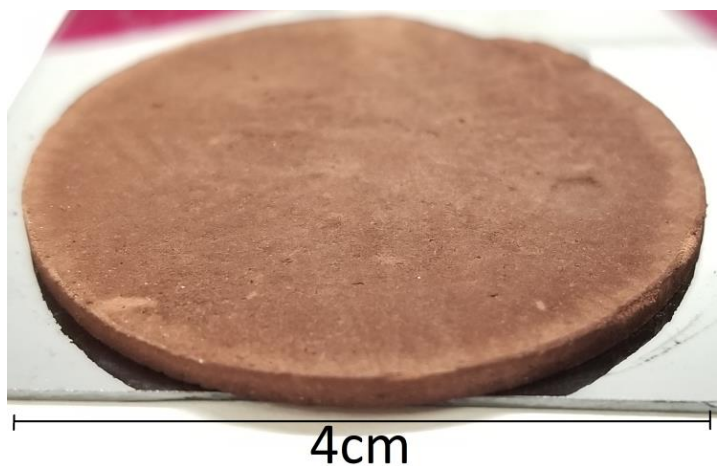

**Figure S7.** Photo of a 1 mm thick, 4 cm diameter copper foam that was incorporated into a respirator cartridge.